\documentclass[aps,prb,twocolumn,amsmath,superscriptaddress,noshowpacs,floatfix]{revtex4-1}
\usepackage{graphicx,bm,epsfig,color,dcolumn,setspace,array,mathrsfs,amsmath,amssymb,textcomp,gensymb,booktabs,url,hyperref,bibentry,natbib}
\begin{document}

\title{Moir\'e magnons in twisted bilayer magnets with collinear order}

\author{Yu-Hang Li}
\email[]{yli@ucr.edu}
\affiliation{Department of Electrical and Computer Engineering, University of California, Riverside, California 92521, USA}

\author{Ran Cheng}
\email[]{rancheng@ucr.edu}
\affiliation{Department of Electrical and Computer Engineering, University of California, Riverside, California 92521, USA}
\affiliation{Department of Physics and Astronomy, University of California, Riverside, California 92521, USA}

\begin{abstract}
We explore the moir\'e magnon bands in twisted bilayer magnets with next-nearest neighboring Dzyaloshinskii-Moriya interactions, assuming that the out-of-plane collinear magnetic order is preserved under weak interlayer coupling.
By calculating the magnonic band structures and the topological Chern numbers for four representative cases, we find that (i) the valley moir\'e bands are extremely flat over a wide range of continuous twist angles;
(ii) the topological Chern numbers of the lowest few flat bands vary significantly with the twist angle; and
(iii) the lowest few topological flat bands in bilayer antiferromagnets entail nontrivial thermal spin transport in the transverse direction.
These properties make twisted bilayer magnets an ideal platform to study the magnonic counterparts of moir\'e electrons, where the statistical distinction between magnons and electrons leads to fundamentally new physical behavior.
\end{abstract}

\maketitle

\section{Introduction}
\label{in}

The experimental discovery of two-dimensional (2D) ferromagnetism~\cite{Huang2017Magneto,Gong2017Intrinsic} spurred a flurry of research, rapidly reshaping the field of 2D materials with an emerging frontier. Since then, 2D magnetic van der Waals materials has become a new and versatile platform to study fundamental physics in reduced dimensions~\cite{Gong2019Two,Burch2018Magnetism,Gibertini2019Magnetic,Tong2018Skyrmions}. A recent highlight in 2D materials is the observation of unconventional superconductivity~\cite{Cao2018Unconventional} and Mott insulator~\cite{Cao2018Correlated} in twisted bilayer graphene where the electronic bands around the Fermi level are nearly flat, magnifying the relative strength of interaction effects.

It has been demonstrated that in insulating magnetic materials, the spin-wave excitations (or magnons) can play the role of electrons in transporting spin angular momenta~\cite{Chumak2015Magnon}. In particular, the Dzyaloshinskii-Moriya interaction (DMI) acts as an effective spin-orbit coupling (SOC) on magnons, enabling the magnonic counterparts of pure spin phenomena usually associated with the electron spin~\cite{Kim2016Realization,Cheng2016Spin,Zyuzin2016Magnon,Owerre2016Topological,Owerre2016A,Owerre2019Topological,Proskurin2017Spin,Loss2017MagTI,Shiomi2017Experimental,Pershoguba2018Dirac,Chen2018Topological,Lee2018StripZigzag,Hayami2016Asymmetric}. At this point, it is tempting to ask if twisting a bilayer magnet can generate any nontrivial magnonic bands and the consequential transport phenomena. If magnon bands become flat, twisted bilayer magnets (TBM) can be exploited to study the magnon-magnon and magnon-phonon interactions, the influences of which have been treated only perturbatively in traditional scenarios. Furthermore, if flat magnon bands are simultaneously topologically non-trivial, exotic thermomagnetic transport can arise. The physics behind flat magnon bands is not a trivial extension of that of electrons because the statistical distinction between magnons and electrons does not just superficially manifest in their thermal population but can also reflect profoundly in the governing equations~\cite{Cheng2016Spin}.

In this paper, we study moir\'e magnons of insulating TBM in honeycomb lattices, where symmetry allows for the next-nearest neighboring DMI in each layer. We assume that the interlayer coupling is of exchange nature and sufficiently weak compared to the intralayer exchange interactions, so the ground state remains in the collinear regime. In a comparative manner, we study four representative ground states of TBM [as illustrated in Fig.~\ref{Schematic}(c)] classified by the relative signs of intralayer and interlayer couplings. The moir\'e band structures will be obtained by plane-wave expansions~\cite{Bistritzer2011Moire}, based on which we will further compute the associated Berry curvatures and topological Chern numbers. Except for the bilayer ferromagnetic (FM) state [case {\bf{i}} in Fig.~\ref{Schematic}(c)], which has the same governing equation as a twisted bilayer transition metal dichalcogenides~\cite{Wu2018Hubbard,Wu2019Topological,Merkl2020,Wang2019Magic,Pan2020Band} or time reversal breaking layered materials~\cite{Lian2020Flat}, all TBMs involving antiferromagnetic (AFM) features [case {\bf{ii}}, {\bf{iii}} and {\bf{iv}} in Fig.~\ref{Schematic}(c)] do not have direct correspondence to electronic systems. Therefore, the appearance of flat magnon bands in TBM reveals a new physical horizon different from the well-established twisted electronic systems.

While magnons are bosonic excitations that preferably populate the high-symmetry $\Gamma$ point, the nontrivial topological properties such as transverse spin transport originate from valley magnons around $K$ and $K^{\prime}$ points, where the bands can reach local minima depending on the DMI. In particular, when two AFM layers are coupled ferromagnetically (such as MnPS$_3$ and its variances~\cite{Wildes1998Spin,Chittari2016Electronic,Chittari2020Carrier}), the lowest few bands in the valleys can be tuned into topologically flat bands at certain twist angle. Even though it is not clear if the flatness of a magnon band is directly linked with its topology, we make a crucial attempt to characterize this subtle relation by calculating the dependence of bandwidth and Chern numbers on the twist angle.

It is worthwhile to distinguish the scope of our work from the findings of two recent papers~\cite{Hejazi2020,Ghader2019Magnon} published during the preparation of our manuscript. Ref.~[\onlinecite{Hejazi2020}] studied the non-collinear ground states of TBM and the twist-induced phase transitions among those states, unravelling the strong interlayer coupling regime. On the other hand, this paper studies the moi\'e magnons on top of \textit{collinear} ground states, which holds in the parameter regime of weak interlayer coupling and strong anisotropy. Ref.~[\onlinecite{Ghader2019Magnon}] focused on the FM ground state [corresponding to our case {\bf{i}} in Fig.~\ref{Schematic}(c)] where the governing equation can be fully mapped to a Schr\"{o}dinger equation describing a twisted bilayer electronic system. In contrast, four representative collinear ground states have been explored in this paper, providing a general and comparative picture of twisted TBM, where we especially identify the unique features originating from the AFM couplings (which can be interlayer, intralayer or both) that has no counterparts in electronic systems.

The rest of this paper is organized as follows. In Sec.~\ref{mf}, we introduce the formalism starting with the model Hamiltonian. In Sec.~\ref{re}, we show the moir\'e band structures and the Berry curvatures associated with the lowest few flat bands. In Sec.~\ref{dis}, we discuss the manipulation of the flatness and topological Chern numbers of the topological flat bands through the twist angle. A brief summary is given in Sec.~\ref{sum}.

\section{Model and formalism}
\label{mf}

Let us consider a magnetic insulator consisting of two honeycomb layers twisted by an angle $\theta$. In the tight-binding limit, this system can be described by the effective Hamiltonian
\begin{align}
&\mathcal{H}=\sum_{n=1,2}\mathcal{H}_n+\mathcal{H}_T,
\label{Full}
\end{align}
with
\begin{subequations}
\label{model}
\begin{align}
&\mathcal{H}_{n}=-J_{1}\sum_{\left<i,j\right>}\bm{S}_{n,i}\cdot\bm{S}_{n,j}+\kappa\sum_{i}(S_{n,i}^z)^2\\\notag
&\qquad\qquad +D_{2}\sum_{\left<\left<i,j\right>\right>}\epsilon_{ij}\bm{\hat{z}}\cdot\bm{S}_{n,i}\times\bm{S}_{n,j},\\
&\mathcal{H}_{T}=-\sum_{i,j}J_{T}(i,j)\bm{S}_{1,i}\cdot\bm{S}_{2,j}.
\end{align}
\end{subequations}
Here, $\mathcal{H}_n$ is the intralayer Hamiltonian for layer $n$ ($n=1,\ 2$ denotes top, bottom layer) and $\mathcal{H}_T$ is the interlayer exchange Hamiltonian with the coupling strength $J_{T}(i,j)$ depending on the distance $r_{ij}$ between site $i$ and site $j$. $J_1$ is the nearest-neighbor exchange interaction ($J_1>0$ for FM and $J_1<0$ for AFM couplings), $D_2$ is the next-nearest neighboring DMI with $\epsilon_{ij}=\pm1$ representing the chirality of atomic bonds, and $\kappa<0$ is the perpendicular easy-axis anisotropy. On a single-layer honeycomb lattice, symmetry forbids nearest-neighboring DMI~\cite{Cheng2016Spin} unless the lattice is strained. By sticking to the regime of weak interlayer coupling, we also omit higher order interlayer interactions beyond $J_T$.

In electronic systems, the tight-binding model is problematic when the bands under consideration is topological nontrivial because constructing well-localized Wannier functions respecting all symmetries becomes impossible~\cite{Thouless1984Wannier,Po2018Fragile,Po2019Faithful}.  Nevertheless, this problem does not even exist in topological magnons as long as the DMI does not destroy the collinear ground state. This is because the real-space basis of spin waves, unlike electronic Bloch waves subject to a Fourier transformation, is infinitely localized on the atomic sites.

\begin{figure}[t]
  \centering
  \includegraphics[width=0.48\textwidth]{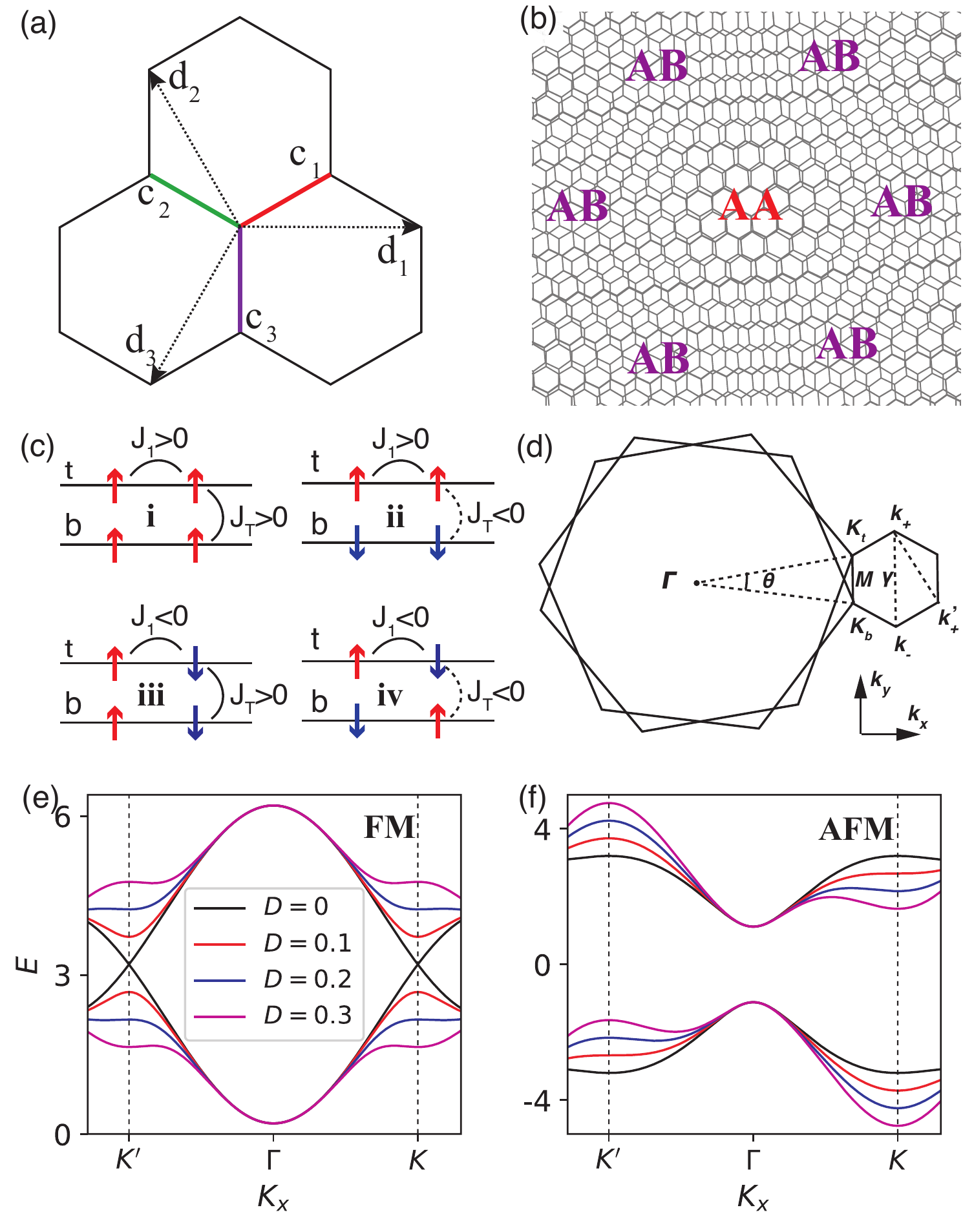}
  \caption{(color online).
  	(a) Schematics for a single-layer honeycomb lattice with $c_{i=1,2,3}$ and $d_{i=1,2,3}$ connecting the nearest and next-nearest neighbors, respectively.
	(b) Moir\'e pattern of a twisted bilayer magnet with alternating $AA$ and $AB$ stacking.
  	(c) Schematic illustration of four different cases of twisted bilayer magnets classified by the signs of $J_1$ and $J_T$.
	(d) The moir\'e BZ formed by overlapping the top and bottom BZs at twist angle $\theta$.
	(e) and (f): band structures of monolayer FM ($J_1>0$) and AFM ($J_1<0$) with different DMIs. The positive (negative) branch in (d) refers to the right-handed (left-handed) magnon excitations.
  	}
\label{Schematic}
\end{figure}

According to the signs of $J_1$ and $J_T$, we classify bilayer magnets into four different cases labeled as {\bf{i}}, {\bf{ii}}, {\bf{iii}} and {\bf{iv}} as illustrated in Fig.~\ref{Schematic}(c). In particular, {\bf{ii}} is representative of layered AFM such as CrI$_3$~\cite{Huang2017Magneto} and {\bf{iii}} is representative of intrinsic AFM with FM interlayer coupling such as MnPS$_3$~\cite{Wildes1998Spin}. At the same incommensurate angle, {\bf{iii}} and {\bf{iv}} exhibit local FM and AFM configurations, respectively, at the AA stacking regions due to the opposite sign of $J_T$. In the following, the moir\'e magnon bands will be calculated by the plane-wave expansion under the single-layer basis, which is valid only when $J_T$ is weak so that the collinear ground state is preserved.

We first solve the single-layer magnon bands. Neglecting magnon-magnon interactions, we adopt the linearized Holstein-Primakoff transformation~\cite{Nolting2009Quantum}  
\begin{equation}
S_{n,i}^{+}\approx\sqrt{2S}a_{n,i},\quad S_{n,i}^{-}\approx\sqrt{2S}a_{n,i}^{\dagger},\quad S_{n,i}^{z}=S-a_{n,i}^{\dagger}a_{n,i},
\label{HPup}
\end{equation}
for up spins and
\begin{equation}
S_{n,i}^{+}\approx\sqrt{2S}b_{n,i}^{\dagger},\quad S_{n,i}^{-}\approx\sqrt{2S}b_{n,i},\quad S_{n,i}^{z}=b_{n,i}^{\dagger}b_{n,i}-S,
\label{HPdown}
\end{equation}
for down spins, where $S_{n,i}^{\pm}=S_{n,i}^x\pm S_{n,i}^y$ and $a_{n,i}^{\dagger}$ ($b_{n,i}^{\dagger}$) is the magnon creation operator for site $i$ on layer $n$ when there is an up (down) spin there. Next we take the Fourier transformation $a_{\bm{k}}=\frac1{\sqrt{N}}\sum_{\bm{k}}e^{-i\bm{k}\cdot\bm{r}_i}a_i$ and $b_{\bm{k}}=\frac1{\sqrt{N}}\sum_{\bm{k}}e^{i\bm{k}\cdot\bm{r}_i}b_i$ to transform $\mathcal{H}_n$ into the momentum space, which in the pseudo-spin representation (\textit{i.e.}, $A$ and $B$ sublattices) takes the form 
\begin{equation}
H_{n}\left(\bm{k}\right)/S=\begin{bmatrix}C+D_{2}g_{n\bm{k}}&&-J_{1}f_{n\bm{k}}\\-J_{1}f^{*}_{n\bm{k}}&&C-D_{2}g_{n\bm{k}}\end{bmatrix},
\label{sham}
\end{equation}
where the zero-point energy has been discarded. If an individual layer is AFM, the pseudospin space is furnished by the Nambu basis $\psi_{\bm{k}}=[a_{\bm{k}}, b^\dagger_{\bm{k}}]^T$ such that $\mathcal{H}_n=\sum_{\bm{k}}\psi_{\bm{k}}^\dagger H_n\psi_{\bm{k}}$. In Eq.~\eqref{sham}, $C=3\left|J_{1}\right|+\left|J_{T}\right|-2\kappa$, $f_{n\bm{k}}=\sum_{i=1}^{3}\exp\left(i\bm{k}\cdot\bm{c}_{n i}\right)$, $g_{n\bm{k}}=2\sum_{i=1}^{3}\left(-1\right)^{i-1}\sin\left(\bm{k}\cdot\bm{d}_{n i}\right)$ where $\bm{c}_{n i}=e^{-i\theta_{n}}\bm{c}_{i}$ and $\bm{d}_{n i}=e^{-i\theta_{n}}\bm{d}_{i}$ with the twist angle of the top (bottom) layer $\theta_{1,2}=\mp\theta/2$. The nearest neighbor links are $\bm{c}_{1}=a_{0}\begin{pmatrix}0,&&-1\end{pmatrix}$ and $\bm{c}_{2,3}=a_{0}\begin{pmatrix}\pm\sqrt{3}/2,&&1/2\end{pmatrix}$, and the next-nearest neighbor links are $\bm{d}_{1}=a_{0}\begin{pmatrix}\sqrt{3},&&0\end{pmatrix}$ and $\bm{d}_{2,3}=a_{0}\begin{pmatrix}-1/2,&&\pm\sqrt{3}/2\end{pmatrix}$. The single-layer band structures obtained by diagonalizing Eq.~\ref{sham} are plotted in Fig.~\ref{Schematic}(e) and (f). For both FM and AFM, the lowest energy appears at the $\Gamma$ point, which is not affected by the DMI. Recent studies have pointed out that even though the $D_2$ interaction originates from the intrinsic symmetry breaking of the honeycomb lattice, which is independent of the Rashba SOC, it does act on the magnons as an effective Rashba SOC thanks to the Holstein-Primakoff transformation~\cite{Kim2016Realization,Cheng2016Spin,Zyuzin2016Magnon,Owerre2016Topological,Owerre2016A,Owerre2019Topological,Proskurin2017Spin,Loss2017MagTI,Shiomi2017Experimental,Pershoguba2018Dirac,Chen2018Topological,Lee2018StripZigzag,Hayami2016Asymmetric}.

Next we turn to the interlayer hopping terms. In the continuum limit~\cite{Bistritzer2011Moire}, the effective interlayer exchange Hamiltonian can be taken as
\begin{equation}
H_T\left(\bm{r}\right)/S=\sum_{i=1}^{3}T\left(\bm{q}_{i}\right)e^{-i\bm{q}_{i}\cdot\bm{r}},
\label{Ham_R}
\end{equation}
where
\begin{equation}
\begin{split}
&T\left(\bm{0}\right)=-J_{T}\begin{bmatrix}0&&1\\1&&0\end{bmatrix},\\
&T\left(\bm{G}_1\right)=-J_{T}\begin{bmatrix}0&&1\\e^{i2\pi/3}&&0\end{bmatrix},\\
&T\left(\bm{G}_1+\bm{G}_2\right)=-J_{T}\begin{bmatrix}0&&1\\e^{-i2\pi/3}&&0\end{bmatrix},
\end{split}
\label{Coupling_Matrices}
\end{equation}
are the interlayer hopping matrices connecting the three nearest neighbors in the momentum space labeled by the reciprocal lattice vectors $\bm{G}_1=K_{\theta}(\frac12,\ \frac{\sqrt{3}}{2})$, $\bm{G}_{2}=K_{\theta}(-1,\ 0)$ with $K_{\theta}=2\cdot 4\pi/\left(3a_{0}\right)\cdot \sin\left(\theta/2\right)$. Similar to the Moir\'{e} electrons~\cite{Tarnopolsky2019Origin}, magnons are essentially accumulated in the $AB$ stacking regions at small twist angles. So, to a good approximation, we set the interlayer hopping for the $AA$ stacking regions to be $t_{AA}=0$ while $t_{AB}=J_T$ for the $AB$ stacking regions, as shown in Fig.~\ref{Schematic}(b). 

We now use the plane-wave expansion to calculate the moir\'e magnon bands of a twisted bilayer magnet~\cite{Bistritzer2011Moire}. Since the interlayer Hamiltonian $H_T(\bm{r})$ is constructed in the real space above, we need to define the real-space intralayer Hamiltonian as well: $H_n(\bm{r})=\int d\bm{k}H_n(\bm{k})e^{-i\bm{k}\cdot\bm{r}}$. Then the total Hamiltonian becomes
\begin{equation}
\mathcal{H}\left(\bm{r}\right)=\begin{bmatrix}H_1\left(\bm{r}\right)&&H_{T}\left(\bm{r}\right)\\H_{T}\left(\bm{r}\right)^{\dagger}&&H_2\left(\bm{r}\right)\end{bmatrix},
\label{Ham_R_matrix}
\end{equation}
from which the moir\'e magnon bands can be solved with a proper truncation in the Brillouin zone (BZ). Different from that of electrons, the eigen-equation for magnons

\begin{equation}
  E\eta\Psi\left(\bm{r}\right)=\mathcal{H}\left(\bm{r}\right)\Psi\left(\bm{r}\right),
  \label{Schrodinger}
\end{equation}
comes with an effective metric
\begin{align}
    \eta=\begin{bmatrix}1&&0\\0&&\mathrm{sgn}\left(J_{T}\right)\end{bmatrix}\otimes\begin{bmatrix}1&&0\\0&&\mathrm{sgn}\left(J_1\right)\end{bmatrix}
\end{align}
that depends on the signs of both interlayer and intralayer couplings. The essential difference in band topology between electrons and magnons lies in this $\eta$ matrix~\cite{Cheng2016Spin}. If one (or both) of $J_T$ and $J_1$ becomes negative (AFM coupling), the eigenvalue problem here will not bear a correspondence in electronic systems, and intuitions acquired from electronic systems no longer apply. Moreover, since symmetry guarantees that the bands of twisted AFM (case {\bf{ii}}, {\bf{iii}}, and {\bf{iv}}) come in pairs of right-handed chirality (positive branch) and left-handed chirality (negative branch) as those in single layer AFM, hereafter we will only show the right-handed ones for simplicity.

To calculate the Berry curvature numerically, we follow the method in Ref.[\onlinecite{fukui2005chern}]. Let the eigenvector in the momentum space be $\Psi\left(\bm{k}\right)$, the Berry curvature can be expressed as 
\begin{align}
\Omega\left(\bm{k}\right)=\frac1{2\pi i}
\ln&\left[U_x(\bm{k})U_y(\bm{k}+\delta k_x)\right. \notag \\
&\left.\times U_x(\bm{k}+\delta k_y)^{-1}U_y(\bm{k})^{-1}\right],
\label{eq:bbb}
\end{align}
where
\begin{equation*}
U_j(\bm{k})=\frac{\langle\Psi(\bm{k})|\eta_{\bm{k}}|\Psi(\bm{k}+k_j)\rangle}{\left|\langle\Psi(\bm{k})|\eta_{\bm{k}}|\Psi(\bm{k}+k_j)\rangle\right|},\quad j=x,y
\end{equation*} 
Here, the factor $\eta_{\bm{k}}=I_N\otimes\eta$ where $I_N$ is an $N\times N$ identity matrix with $N$ the dimension of the Hamiltonian after truncation.
The Chern number can then be obtained straightforwardly by integrating the Berry curvature over the first BZ
\begin{align}
C=\frac{1}{2\pi}\int{d^2k\Omega\left(\bm{k}\right)}.
\end{align}
In our calculation, the plane-wave expansion is truncated at the crystal momentum $k_{t}=4K_{\theta}$ if $\theta>1^{\degree}$ and $k_{t}=4K_{\theta}/\theta$ otherwise -- a strategy widely adopted in the study of Moir\'{e} electrons~\cite{Bistritzer2011Moire}. We adopt the scaling convention that the lattice constant $a_{0}=1$ and $\left|J_1\right|=1$. Other parameters are taken to be $S=1$, $\kappa=-0.05$, and $\left|J_{T}\right|=0.1$ unless otherwise specified.

\section{Band Structures, Topology, and Flatness}
\label{re}

\subsection{$\Gamma$ point magnons}
\label{gpoint}

Magnons are bosons that preferably occupy lower-energy states. Therefore, at low temperatures, magnons are mostly populated in the vicinity of the $\Gamma$ point. However, because of high symmetry, the moir\'e bands near new center of the shrunk moir\'e BZ--$\gamma$ point--do not take on essential new features beyond what already exist near the single-layer $\Gamma$-point, thus impatient readers can skip to the next subsection directly.

\begin{figure}[t]
  \centering
  \includegraphics[width=0.48\textwidth]{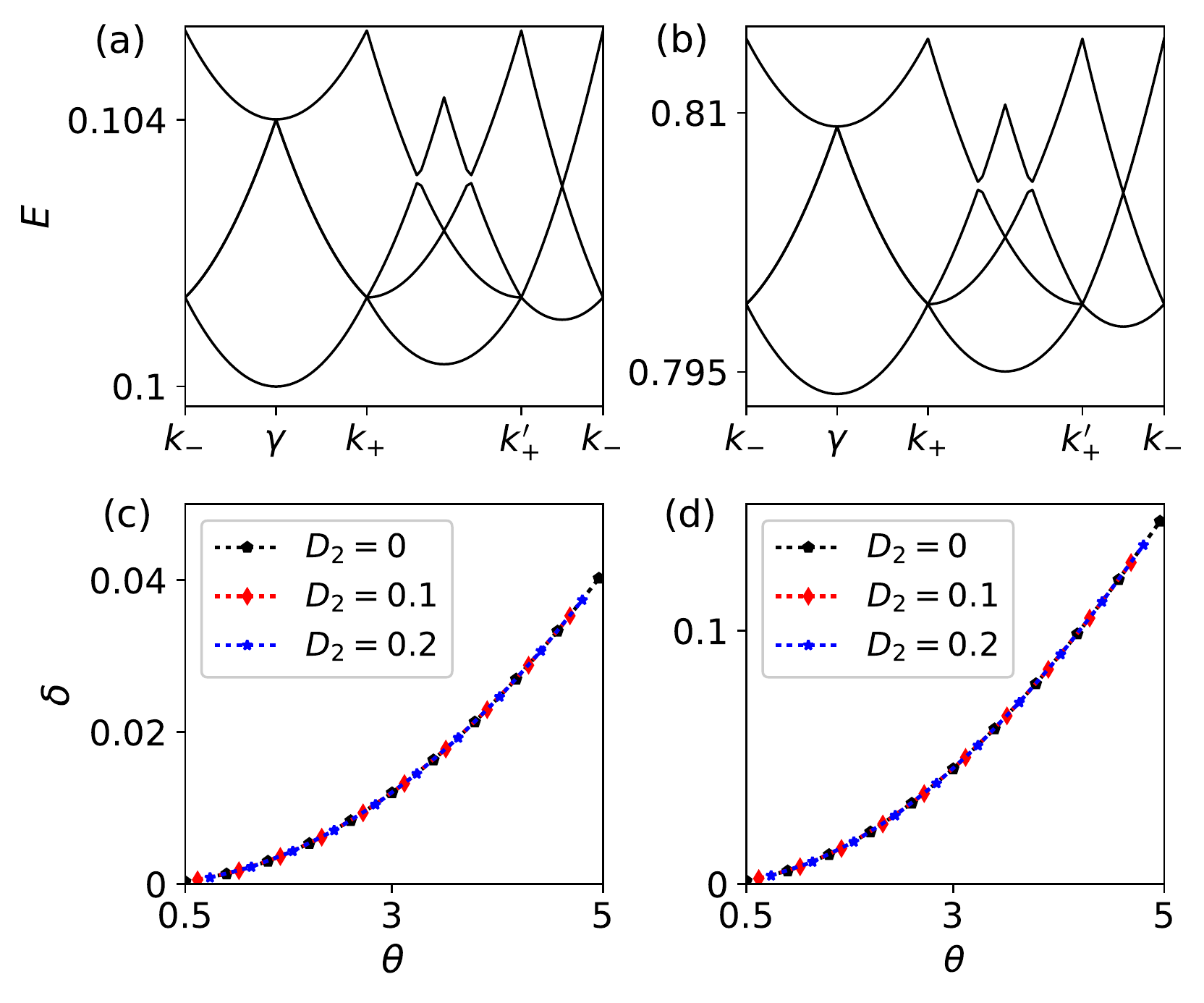}
  \caption{(color online).
  	(a) and (b): the lowest moir\'e bands shrunk from the $\Gamma$ point of the FM (case {\bf{i}}) and AFM (case {\bf{iii}}) single layers at $\theta=1^{\degree}$. In both cases, $J_{T}=0.1$ and $D_{2}=0$. (c) and (d): the widths of the lowest band as a function of $\theta$ for different values of DMI.
  	}
\label{G_point}
\end{figure}

Nevertheless, for theoretical completeness, it is instructive to first study the comparably trivial $\Gamma$ point. The commensurate condition requires that the $\Gamma$ points of the top and bottom layers coincide with the new $\gamma$ point of the moir\'e BZ up to integers of reciprocal lattice vector~\cite{Lopes2012Continuum} as illustrated in Fig.~\ref{Schematic}(d). Consequently, twisting only introduces a phase difference between the Hamiltonians of the top and bottom layers within only the nearest-neighbor coupling in the momentum space. Meanwhile, the interlayer Hamiltonian reduces to  $H_T\left(\bm{r}\right)=T\left(\bm{0}\right)=-J_{T}\left[\begin{matrix}1&&0\\0&&1\end{matrix}\right]$. Then the total Hamiltonian becomes
\begin{equation}
\mathcal{H}=\left[\begin{matrix}H_1\left(\bm{k}\right)&&T\left(\bm{0}\right)\\T\left(\bm{0}\right)^{\dagger}&&H_2\left(\bm{k}\right) \end{matrix}\right],
\label{Ham_K}
\end{equation}
where $\bm{k}$ is restricted to the first moir\'e BZ and the band structure can be solved by directly diagonalizing this Hamiltonian. Near the $\Gamma$ point, single-layer bands shrink into the moir\'e BZ with minor deformation due to the weak interlayer coupling. Fig.~{\ref{G_point}}(a) and (b) plot the lowest few bands for cases {\bf{i}} and {\bf{iii}} at twist angle $\theta=1^{\degree}$, where the only visible difference is the zero energy gap at the $\gamma$ point. Those bands are quadratic, topologically trivial and independent of the DMI. Moreover, varying the twist angle does not lead to flat bands. Fig.~\ref{G_point}(c) and (d) plot the width of the lowest band as a function of $\theta$ at different DMIs, where the reduction of the bandwidth with a decreasing $\theta$ is simply due to the shrinking of the moir\'e BZ instead of any twist-induced qualitative change.

To confirm that we have not missed any essential physics due to crude approximation, we have refined the above calculation by both including higher order terms in the interlayer coupling (in the $\bm{k}$ space) and truncating the plane-wave expansion at larger momenta. However, by doing so, we do not observe any visible change with the same resolution in Fig.~\ref{G_point}.

\subsection{$K$ point magnons}
\label{kpoint}

Even though single-layer magnons are mostly populated near the $\Gamma$ point, non-trivial transport properties are attributed to magnons from the $K$ and $K'$-valleys~\cite{Kim2016Realization,Cheng2016Spin,Zyuzin2016Magnon,Owerre2016Topological,Owerre2016A}. As shown in Fig.~\ref{Schematic}(e) and (f), a substantial DMI can even yield the energy of valley magnons comparable to that of $\Gamma$ magnons. Therefore, with an increasing temperature, valley magnons become more and more important in determining spin transport. We expect similar outcomes for the mori\'e magnons because the mori\'e bands are calculated based on single-layer Hamiltonians in the presence of interlayer coupling.

\begin{figure*}[t]
  \centering
  \includegraphics[width=1.0\textwidth]{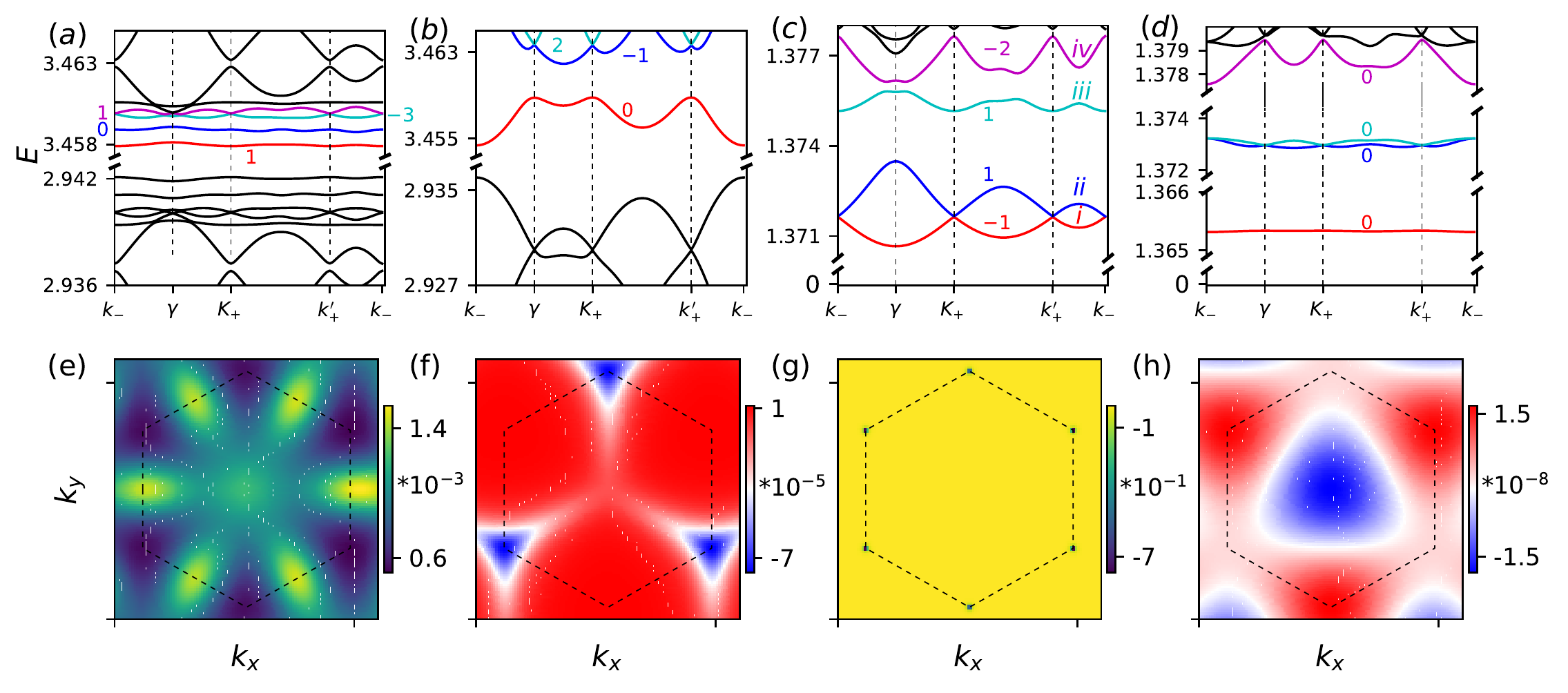}
  \caption{(color online).
  	(a), (b), (c) and (d) are the moir\'e bands of valley magnons of the four cases defined in Fig.~\ref{Schematic}(c). The band Chern numbers are labeled on the top of each band.  Note that the band Chern number is not well defined if two bands intersect with each other. The DMI is taken to be 0.05 in (a) and (b), and 0.35 in (c) and (d). The twist angle is $\theta=1.0^{\degree}$. (e), (f), (g) and (h) are the corresponding Berry curvatures of the lowest band colored red.
  	}
\label{Band_Curvature}
\end{figure*}

To study magnons folded from both the $\Gamma$ and $K$ points uniformly, we shift the $\gamma$ point of the shrunk moir\'e BZ to the origin by adding a twist-angle dependent constant $\bm{k}_{0}=\left\{4\pi\cdot\left(\cos{\theta/2}+\sqrt{3}\sin{\theta/2}\right)/\left(3\sqrt{3}a_{0}\right),\ 0\right\}$ to the crystal momentum, defining $\tilde{H}_n\left(\bm{r}\right)=\int d\bm{k}H_n(\bm{k}+\bm{k}_0)e^{-i\bm{k}\cdot\bm{r}}$ with $\bm{k}$ restricted to the shrunk moir\'e BZ. Hence, the real-space Hamiltonian becomes
\begin{equation}
\mathcal{H}\left(\bm{r}\right)=\begin{bmatrix}\tilde{H}_1\left(\bm{r}\right)&&H_{T}\left(\bm{r}\right)\\H_{T}\left(\bm{r}\right)^{\dagger}&&\tilde{H}_2\left(\bm{r}\right)\end{bmatrix},
\label{Ham_K_Kpoint}
\end{equation}
where $\bm{k}$ is restricted to the shrunk moir\'e BZ. Fig.~\ref{Band_Curvature} plots the moir\'e bands and the Berry curvatures associated with the lowest valley band colored red at $\theta=1^{\degree}$ for the four different cases illustrated in Fig.~\ref{Schematic}(c).

In the bilayer FM case shown in Fig.~\ref{Band_Curvature}(a), the bands are symmetric about the plane $E_c=3\left|J_{1}\right|+\left|J_{T}\right|-2\kappa=3.2$; the Chern numbers of the bands above $E_c$ are opposite to those below $E_c$. Regarding the mathematical similarity between FM magnons and Haldane's electrons on the single-layer honeycomb lattice~\cite{Owerre2016A,Owerre2016Topological,Kim2016Realization}, the eigenvalue problem in the twisted bilayer FM can be viewed as a magnonic version of twisted bilayer transition metal dichalcogenides~\cite{Merkl2020,Wang2019Magic}, where the Rashba SOC plays a crucial role in determining the band structures. Therefore, it is straightforward to understand why the two bands mostly close to $E_c$ are flat basing on intuitions of Moir\'{e} electrons. A recent study strongly suggests that $D_2$ is appreciably strong~\cite{Chen2018Topological}, which under the analogy to Haldane's electrons, corresponds to a much stronger effective Rasbha SOC acting on the magnons.

In the other three cases plotted in Fig.~\ref{Band_Curvature}(b), (c) and (d), AFM characteristics manifest in one way or another as the metric $\eta$ appearing in Eq.~\ref{Schrodinger} involves the $-1$ factors. But they all share the common feature that magnon bands come in pairs of right-handed and left-handed chiralities, which mirror each other under certain symmetry transformations. This trait of symmetry has been discussed in detail for a single-layer honeycomb AFM~\cite{Cheng2016Spin,Zyuzin2016Magnon}, but it becomes much less obvious when turning to the twisted bilayers. We believe that the reason is that the plane wave expansion is performed on the single-layer basis, which holds only for weak $J_T$.

Similar to the bilayer FM case, in Fig.~\ref{Band_Curvature}(b), (c) and (d) the widths of the first few bands (colored) above the lowest energy point of single-layer valley magnons turn out to be negligible compared to the band gaps. It is noticeable that, when expressed in absolute values, they are even flatter than the electronic flat bands in twisted bilayer graphene. Unlike the bilayer FM case, however, they have no direct correspondence to the electronic systems because of the metric $\eta$, thus the flat bands in twisted AFMs are not superficial extensions of any flat-band electronic systems.

The topological properties embedded in those flat bands are indicated by the Chern numbers marked in Fig.~\ref{Band_Curvature}. While bands symmetrically distributed around $E_c=3.2$ in Fig.~\ref{Band_Curvature}(a) have opposite Chern numbers, the effective symmetry plane in twisted bilayer AFMs is $E=0$, which separates the right-handed sector from the left-handed sector~\footnote{The eigenfrequency $\omega$ can be both positive and negative, depending on the chirality of magnons. But the energy is always positive. So here, $E$ should be understood as $\hbar\omega$.}. Accordingly, we find that the Chern numbers of the right-handed and the left-handed modes differ by a minus sign, allowing pure spin transport to be elaborated in the next section. Except for the case of Fig.~\ref{Band_Curvature}(b)--layered AFM, the energetically relevant moir\'e magnon bands are either flat or topologically non-trivial, or both. In particular, the lowest flat band shown in Fig.~\ref{Band_Curvature}(c) is simultaneously topologically non-trivial, which is amenable to thermally-induced pure spin transport. Even though the Chern numbers of the lowest four bands are all zero in Fig.~\ref{Band_Curvature}(d) at $\theta=1.0^{\degree}$, they can be tuned by varying $\theta$ as will be shown in Fig.~\ref{cnum_Kxy} later. Moreover, the Berry curvature exhibits a six-fold (three-fold) rotational symmetry if the interlayer exchange interaction $J_T$ is FM (AFM), which can be attributed to the $\eta$ factor unique to bosonic systems.

In deriving the above results, we find that the plane-wave expansion is good only if the local band dispersion around the (single-layer) $K$ points is above a certain minimum. This sets an upper limit of DMI about $0.1$ for case {\bf{i}} and {\bf{ii}} and a lower limit of DMI about $0.3$ for case {\bf{iii}} and {\bf{iv}}, as the twist-angle varies from $0.5^{\degree}$ to $5^{\degree}$.

\section{Manipulating Flatness and Topology By Twisting}
\label{dis}

In this section we discuss how the width and topology of the lowest few flat bands of valley magnons (colored in Fig.~\ref{Band_Curvature}) depend on system parameters and the resulting effects in the transverse transport.

\begin{figure}[t]
  \centering
  \includegraphics[width=0.5\textwidth]{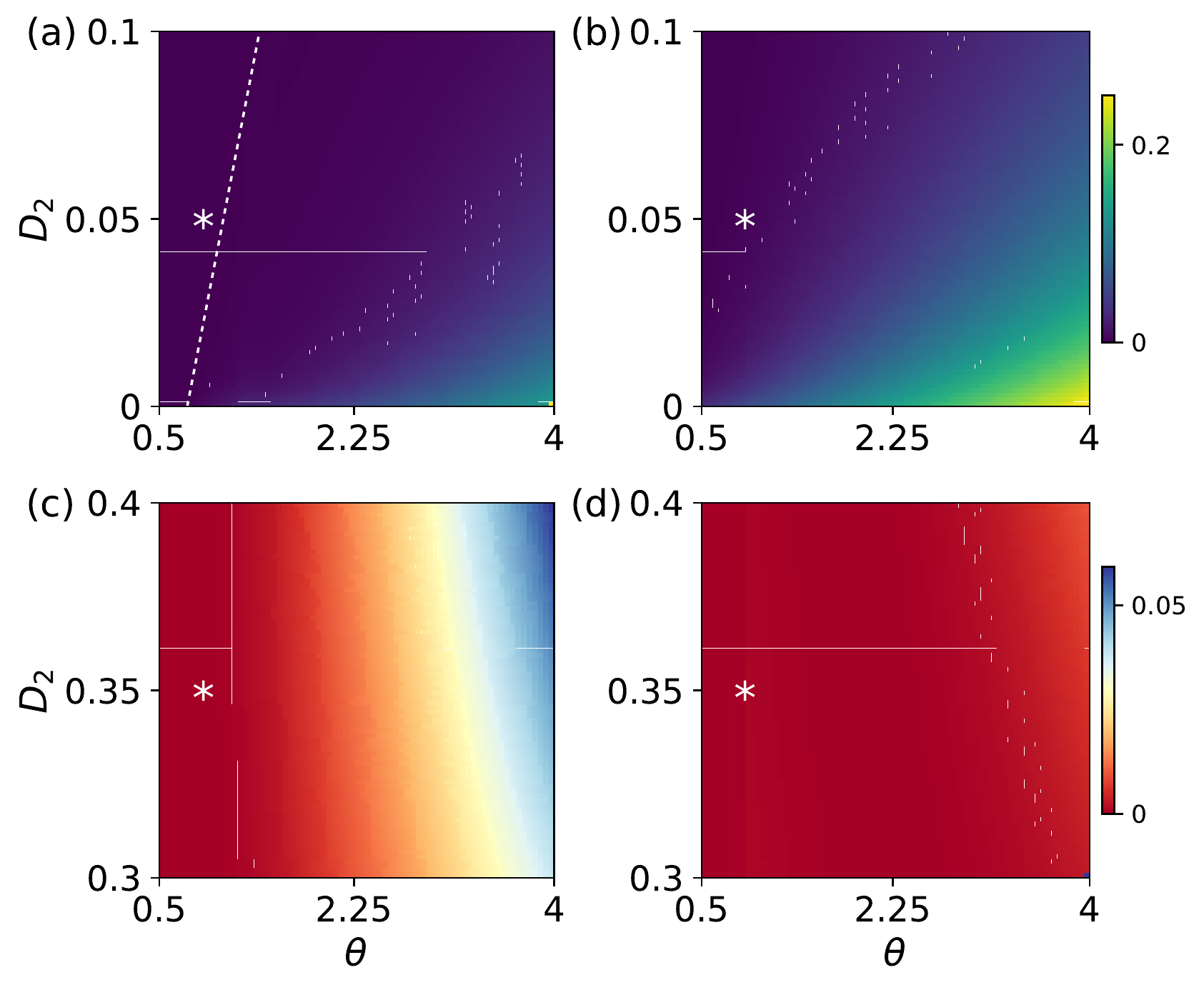}
  \caption{(color online).
  	Bandwidth $\delta$ of the lowest flat bands (see red curves in Fig.~\ref{Band_Curvature}) as a function of the twist angle $\theta$ and the DMI $D_{2}$ for $|J_{T}|=0.1$. The white dashed line in (a) indicates the local minima of $\delta$. The white star marks where parameters in Fig.~\ref{Band_Curvature} are chosen: $\theta=1^{\degree}$ and $D_{2}=0.05$ ($0.35$) for cases {\bf{i}} and {\bf{ii}} ({\bf{iii}} and {\bf{iv}}).
  	}
\label{D2}
\end{figure}

Fig.~\ref{D2} plots the width $\delta$ of the lowest valley magnon band as a function of the twist angle $\theta$ and the DMI $D_2$ for the four different cases, respectively. When the single-layer ground state is FM (AFM), the range of $D_2$ is restricted to be $0$ to $0.1$ ($0.3$ to $0.4$) to ensure the validity of the continuum model. We notice that this band is kept flat over a wide continuum of $\theta$ rather than showing up at discrete and sharp ``magic angles". Of the four cases, only the bilayer FM (case {\bf{i}}) shown in Fig.~\ref{D2}(a) is analogous to a twisted transition metal dichalcogenide~\cite{Merkl2020,Wang2019Magic}, whereas all the other cases do not have direct correspondence to any electronic systems due to the $\eta$ matrix in Eq.~\ref{Schrodinger}. While we can still find the local minima of $\delta$ as marked by the dashed line, it no longer represents where the band suddenly turns flat, which is almost invisible for any infinitesimal $D_2$. $\delta$ shows almost a monotonic dependence on $\theta$. However, the sharp pattern of discrete ``magic angles" where $\delta$ suddenly turns flat can be retrieved in Fig.~\ref{D2}(a) (case {\bf{i}}) when $D_2$ is exactly zero, where both the Hamiltonian and the governing equation reduce to those of a twisted bilayer graphene. Here, an infinitesimal $D_2$ opens a topological non-trivial gap at the $K$ and $K'$ points in the single-layer FM band~\cite{Kim2016Realization,Owerre2016Topological}, which amounts to a discontinuous change of the bases in diagonalizing Eq.~\ref{Ham_K_Kpoint}. The above finding indicates a profound fact beyond the magnonic mori\'e systems: even in twisted bilayer electronic systems, including the Rashba SOC may remove the pattern of ``magic angles" and introduce a wide continuum of flat bands~\cite{Wang2019Magic}.

\begin{figure}[t]
  \centering
  \includegraphics[width=0.5\textwidth]{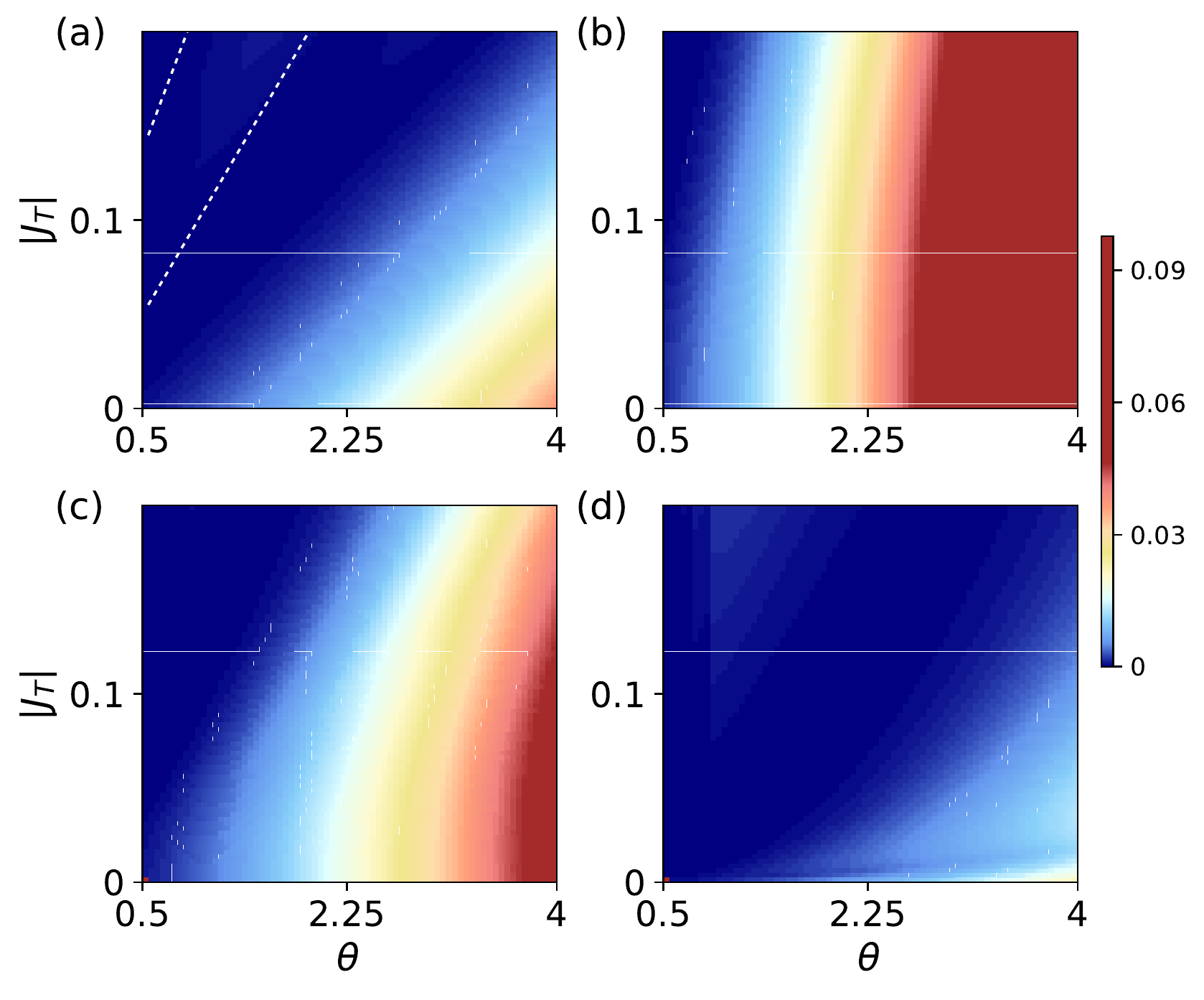}
  \caption{(color online).
  Bandwidth $\delta$ of the lowest flat band (red curve in Fig.~\ref{Band_Curvature}) as a function of the twist angle $\theta$ and the interlayer interaction $J_{T}$ at $D_2=0.05$ ($0.35$) for cases {\bf{i}} and {\bf{ii}} ({\bf{iii}} and {\bf{iv}}). The white dashed lines in (a) denote the local minima of $\delta$.
  	}
\label{Jt}
\end{figure}

Fig.~\ref{Jt} further shows the dependence of $\delta$ on the interlayer exchange interaction $J_{T}$. For a given twist angle, the bandwidth decreases with an increasing $J_{T}$ in all cases, suggesting that the interlayer coupling enhances the lowest flat band. In Fig.~\ref{Jt}(a), the white dashed lines mark the local minima of $\delta$. Similar to Fig.~\ref{D2}, the bandwidth $\delta$ does not suddenly turn flat on these lines but remain essentially flat even away from these local minima, persisting over a wide range of angles. In addition, We notice that the bandwidth of case {\bf{iv}} [Fig.~\ref{Jt}(d)] is much smaller than other cases, which is consistent with and in fact incorporates what we have shown in Fig.~\ref{Band_Curvature}(d).

Next we turn to the topological property and the resulting transverse magnon transport. As has been established in single-layer magnonic systems, under a temperature gradient $\nabla T$, the band $m$ with a Berry curvature $\Omega_m(\bm{k})$ contributes to the total magnon flow with a magnon Hall current density as~\cite{Matsumoto2011Theoretical,Xiao2006Berry}
\begin{equation}
\begin{split}
\bm{J}_m=&\frac{k_{B}}{h}\hat{\bm{z}}\times\nabla{T} \int{\frac{d^2k}{2\pi}\Omega_{m}\left(\bm{k}\right)}\left\{\right.\rho_{m}\left(\bm{k}\right)\ln{\rho_{m}\left(\bm{k}\right)}\\
&-\left[1+\rho_{m}\left(\bm{k}\right)\right]\ln{\left[1+\rho_{m}\left(\bm{k}\right)\right]}\left.\right\}, 
\end{split}
\end{equation}  
where $k_B$ is the Boltzmann constant, $h$ is the Planck constant, and $\rho_m(\bm{k})=1/[e^{E_m(\bm{k})/k_{B}T}-1]$ is the Bose-Einstein distribution function. If band $m$ is flat (\textit{i.e.}, $E_m(\bm{k})$ is approximately a constant), $\rho_m$ will be approximately a constant throughout the moir\'e BZ. As a result, $\bm{J}_m$ can be simplified into $\bm{J}_m\approx\frac{k_B}{h}\hat{\bm{z}}\times\nabla{T} f_mC_m$ where $C_m$ is the Chern number and $f_m=\rho_m\ln{\rho_m}-(1+\rho_m)\ln{(1+\rho_m)}$ is a constant depending only on temperature. When the lowest few bands are all flat, the total magnon Hall current density has a dominant contribution proportional to $\sum_mf_mC_m$. This simplification due to band flatness allows us to analyze the transverse transport in terms of topological Chern numbers (at least qualitatively) even though magnons are statistically distinct from electrons.

Case {\bf{i}} is quite straightforward. Even though we cannot obtain the dispersion relations and Chern numbers of all bands due to the limit of plane-wave expansion, we can claim a finite magnon Hall effect based on the acquired information. As discussed above, the bands in Fig.~\ref{Band_Curvature}(a) are symmetrically distributed around $E_c$ and exhibit opposite Chern numbers for $E>E_c$ and $E<E_c$. On the other hand, the thermal distribution $\rho_m$ does not respect such symmetry. It is highly skewed around $E_c$. Therefore, the magnon Hall currents arising from different bands, especially the contribution of the first few flat bands $\sum_mf_mC_m$, do not cancel.

\begin{figure}[t]
  \centering
  \includegraphics[width=0.5\textwidth]{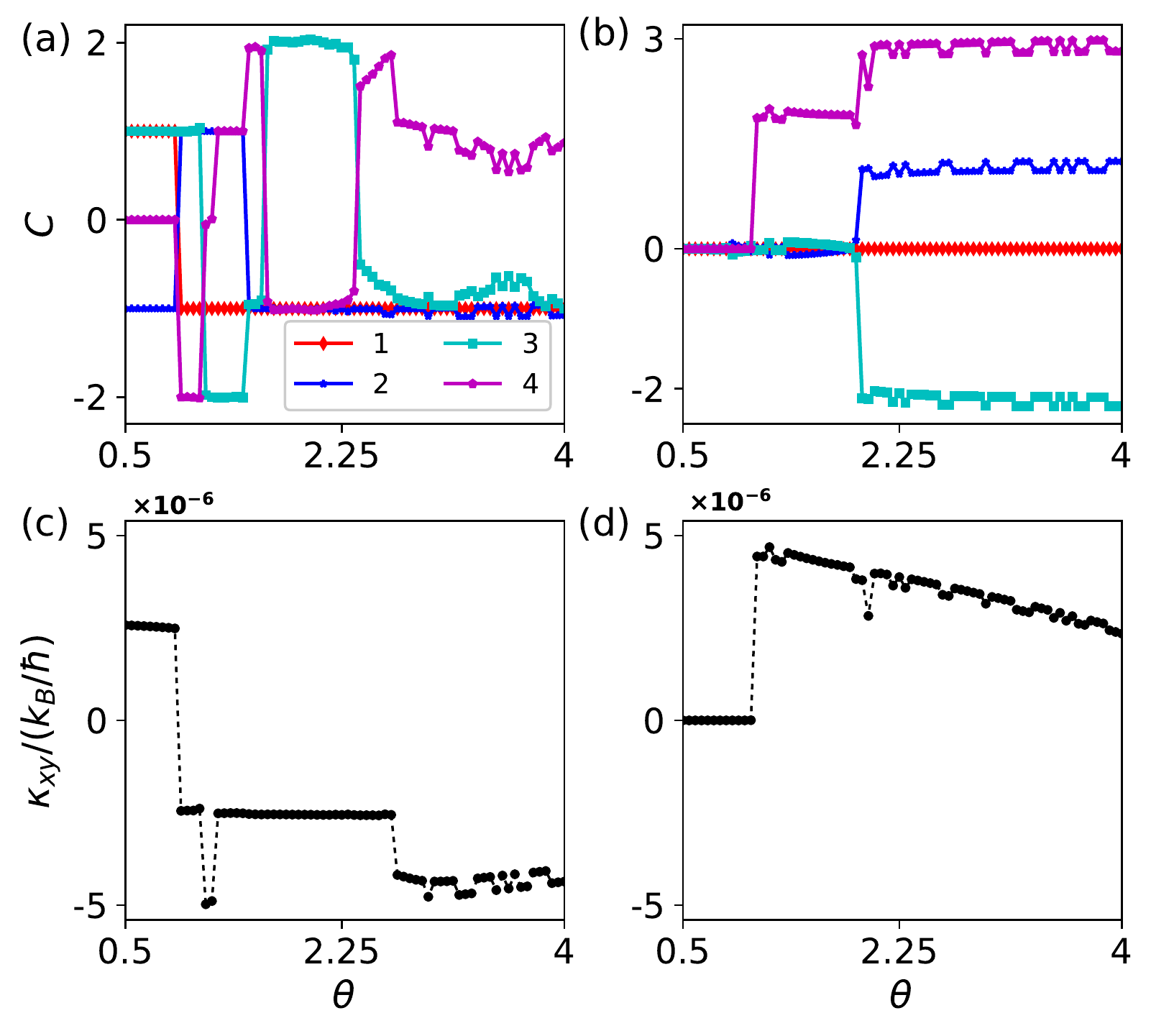}
  \caption{(color online).
    (a) and (b) are the Chern numbers of the lowest four bands depicted in Fig.~\ref{Band_Curvature}(c) and (d) as functions of the twist angle with the same coloring. (c) and (d) are the total contribution of these four bands to the SNE coefficient at $T=0.1$. Other parameters are taken to be the same as those in Fig.~\ref{Band_Curvature}.
  	}
\label{cnum_Kxy}
\end{figure}

The most interesting cases are {\bf{iii}} and {\bf{iv}}, where the lowest few bands are topological flat bands so that they are energetically relevant to the transverse spin transport. In Fig.~\ref{cnum_Kxy}(a) and (b), we plot the Chern numbers of the lowest four topological flat bands as functions of $\theta$ with color schemes matching those in Fig.~\ref{Band_Curvature}(c) and (d). It is interesting to see that even though Fig.~\ref{Band_Curvature}(d) (case {\bf{iv}}) exhibits all zero Chern numbers at $\theta=1^{\degree}$, changing $\theta$ can induce nonzero Chern numbers, resulting in many topologically distinct phases within the range of $0.5^{\degree}<\theta<4^{\degree}$. Besides, the right-handed and left-handed magnons always have opposite Chern numbers regardless of the twisting angle, which establishes a direct connection between non-zero Chern numbers and the spin Nernst effect (SNE). Therefore, twisting becomes an effective operation to manipulate the band topology, hence the transverse spin transport in bilayer AFMs. Although this does not reflect the overall magnitude of SNE because higher bands are neglected, it reveals that the dominant contribution of SNE is indeed controllable via twisting.

Fig.~\ref{cnum_Kxy}(c) and (d) show the contributions of the lowest four topological flat bands to the SNE coefficient $\kappa_{xy}$, defined as $\kappa_{xy}=\hbar(J^{\uparrow}-J^{\downarrow})/(\hat{z}\times\nabla{T})$ with $J^{\uparrow}(J_{\downarrow})$ the Hall current originated from left (right)-handed magnon band, for the cases of {\bf{iii}} and {\bf{iv}} in Fig.~\ref{cnum_Kxy}(a) and (b), respectively.  In particular, $\kappa_{xy}$ can even flip sign by varying $\theta$ as shown in Fig.~\ref{cnum_Kxy}(c). While a more accurate and complete investigation of higher bands is needed to determine quantitatively the SNE in twisted bilayer magnonic systems, Fig.~\ref{cnum_Kxy} provides a first-step exploration into AFM moir\'e magnons in TBM, the dynamics of which, unlike twisted bilayer FM, have no direct correspondence to any electronic systems. In all four cases of TBM studied above, twisting proves to be a useful tuning knob of both the band flatness and the band topology. The subtle relation between the two seemingly different properties will be investigated in future studies.

\section{Summary}
\label{sum}
In summary, we have studied the moir\'e magnons in twisted bilayer magnets for four representative cases. By using the plane-wave expansion, we demonstrated that the moir\'e magnon bands shrunk from the $\Gamma$ point of the single-layer magnet are neither flat nor topologically nontrivial. In striking contrast, magnon bands shrunk from the $K$ point become topological flat bands in the moir\'e BZ over a wide range of twist angles, which is robust against the interlayer exchange interaction as long as the collinear ground state is preserved. Those topological flat bands can generate  transverse magnon transport that is controllable by the twist angle, opening an intriguing playground for topological Moir\'{e} magnons.

Finally, we remark that the rapid development in 2D magnets is approaching a similar level of understanding as 2D electronic materials~\cite{Merkl2020}. In particular, monolayer magnet of long-range collinear order has been realized experimentally in varies magnetic materials. Considering that Moir\'{e} electronic bilayers can be fabricated by ``tear-and-stack" technique using monolayers, we expect a similar technique be devised for 2D magnets in the perceivable future.

\section*{ACKNOWLEDGMENTS}
\label{ack}
 This work is supported by the startup fund of UC-Riverside. The authors acknowledge useful discussions with Y. Zhang.

\bibliography{citation}

\end{document}